# Distributed Joint Source-Channel Coding for Functions over a Multiple Access Channel

R Rajesh and Vinod Sharma
Dept. of Electrical Communication Engg.,
Indian Institute of Science, Bangalore, India
Email: rajesh@pal.ece.iisc.ernet.in,vinod@ece.iisc.ernet.in

*Abstract*— In this paper we provide sufficient conditions for lossy transmission of functions of correlated data over a multiple access channel (MAC). The conditions obtained can be shown as generalized version of Yamamoto's result [28]. We also obtain efficient joint source-channel coding schemes for transmission of discrete and continuous alphabet sources to recover the function values.

## I. INTRODUCTION AND SURVEY

Sensor networks are used in a wide variety of applications due to their ability to operate in environments where human penetration is not possible. One killer application of such networks is environmental monitoring. Correlated sources are a common scenario in densely packed sensor networks. The nodes in a sensor network may be arranged in a hierarchical fashion where neighboring nodes first transmit to a cluster head and then the cluster heads transmit to the fusion center, forming a network of multiple access channels (MACs) ([23]). Also the cluster heads often need to send only a function of the sources to the fusion center. Typical functions include average, maximum, minimum, median or a binary function of the sensor readings ([12], [9]). For example sending the maximum of observations to the fusion center has applications in sensor networks used for fire detection in a building or oil spillage in sea.

In such a set up the sensor nodes can compress the data sent to the cluster head exploiting the correlation in the data and also the structure of the function to be computed at the cluster head. Depending upon the function, exploiting the structure of the function can substantially reduce the data rate for transmission. Since source-channel separation may not hold in this case even for independent sources ([18]), one needs joint source-channel codes to transmit over the MAC. We provide joint source-channel codes that can be used for transmission of functions and give an interpretation of the auxiliary random variables arising in the coding schemes for a general class of functions.

In the following we survey the related literature. The seminal paper, Cover, El Gamal and Salehi [5] provided sufficient conditions for transmitting losslessly correlated observations over a MAC. They also show that unlike for independent sources, the source-channel separation does not hold for this system. The results of [5] have been extended in [24] to the system with continuous alphabets, lossy transmission and side information.

Korner and Marton [14] found the rate region for sending binary sum of two correlated uniform binary sources in a distributed manner. They show that $R_1 > H(f(U_1, U_2))$ and $R_2 > H(f(U_1, U_2))$ is achievable where $R_1$ and $R_2$ are the rates of transmission, $(U_1, U_2)$ are the sources and $f(U_1, U_2) = U_1 \oplus U_2$, $\oplus$ denoting exclusive-OR function. Their proof relies on linear codes as opposed to the more prevalent random codes. Yamamato [28] found the rate-distortion function for sending a source $X$ to a decoder that has $Y$ as side-information and must estimate $f(X, Y)$ to a given distortion $D$. Feng, Effros, and Savari [11] extended this work to a system with noisy observations.

Reference [16] uses the Korner-Marton scheme for finding achievable rate region for distributed source coding where the decoder is interested in lossy reconstruction of an arbitrary function. Linear codes were also used in [18] for joint source-channel codes that are optimal for a class of functions and channels. They show that for reliably constructing a function over a multiple access channel source-channel separation does not hold even when the sources are independent. Exact rate region is obtained when the MAC computes a linear function over a finite field of its inputs followed by a symmetric discrete memoryless channel. Achievability conditions for a general MAC are given using systematic codes.

Orlitsky and Roche [19] used a graph theoretic formulation to find the required rate for sending $X$ to a decoder with side information $Y$ for reliable computation of $f(X, Y)$. They show that the required rate for the above problem is $H_G(X|Y)$ where $H_G(X|Y)$ denotes the conditional entropy of the characteristic graph. The characteristic graph is defined by Witsenhausen [27]. An excellent survey of graph entropy is given in [25].

Vishal et al. [8] established that the minimum conditional entropy coloring of the OR product of the characteristic graph is asymptotically equal to the conditional entropy. Hence they provide a technique to separate out the functional coding and correlation coding by first coloring the graph and then applying Slepian-Wolf coding. It is also possible to extend this technique to lossy transmission. The minimum entropy coloring from a computational view point and from an algorithmic and

This work was partially supported by the DRDO-IISc program on Advanced Research in Mathematical Engineering.

heuristics view point are given in [2] and [3].

Reference [9] considers the distributed encoding of functions of correlated sources and obtains a multi-letter condition for the problem. It is shown that under certain conditions known as the 'zig-zag' condition it is possible to characterize the region by means of first coloring the graphs of $U_1$ and $U_2$ separately and then using Slepian-Wolf codes for the colors.

The paper makes the following contributions. We provide joint source-channel coding schemes for computation of general functions over a general MAC. The sources can be correlated and the source/channel alphabets can be discrete or continuous. The transmission can be lossy or lossless. Also side information may be present. For this set up we have developed general joint source-channel coding techniques. The existing literature on function transmission either studies only source coding or discrete sources over a discrete alphabet MAC. Also it mostly considers simple or specific functions. We show previous results as special cases of our results. The emphasis is on developing general techniques.

The rest of the paper is organized as follows. Section II gives the sufficient conditions for lossy transmission of functions over a MAC. Special cases are given in Section III and Section IV considers joint source-channel coding using graph coloring. Section V gives an example to motivate joint source-channel codes using graph coloring. Section VI considers the joint source-channel coding for continuous alphabet sources and/or channels. Section VII concludes the paper. The proof of Theorem 1 is given in Appendix.

## II. TRANSMISSION OF FUNCTIONS OF SOURCES OVER A MAC

In this section we consider the transmission of functions of memoryless dependent sources, through a memoryless multiple access channel in the presence of side information. The sources and/or the channel input/output alphabets can be discrete or continuous.

We consider two sources $(U_1, U_2)$ and side information random variables $Z_1, Z_2, Z$ with a known joint distribution $F(u_1, u_2, z_1, z_2, z)$. Side information $Z_i$ is available to encoder $i$, $i = 1, 2$ and the decoder has side information $Z$. The random vector sequence $\{(U_{1n}, U_{2n}, Z_{1n}, Z_{2n}, Z_n), n \geq 1\}$ formed from the source outputs and the side information with distribution $F$ is $iid$ in time. The sources transmit their codewords $X_{in}$'s to a single decoder through a memoryless MAC. The channel output $Y$ has distribution $p(y|x_1, x_2)$ if $x_1$ and $x_2$ are transmitted at that time. The decoder receives $Y_n$ and also has access to the side information $Z_n$. The encoders at the two users do not communicate with each other except via the side information. The decoder uses the channel outputs and its side information to estimate a function $G = f(U_1, U_2)$ of sensor observations as $\hat{G}$. It is of interest to find encoders and a decoder such that $\{U_{1n}, U_{2n}, n \geq 1\}$ can be transmitted over the given MAC with $E[d(G, \hat{G})] \leq D$ where $d$ is a non-negative distortion measure and $D$ is the given distortion constraint. We will assume that $d(a, a') = 0$ if and only if $a = a'$. We also assume that either $d$ is upper bounded by $d_{max} < \infty$ or there exist $(u_1^*, u_2^*)$ such that $E[d(f(U_1, U_2), f(u_1^*, u_2^*))] < \infty$. Source-channel separation does not hold in this case.

*Definition*: The source $(U_1^n, U_2^n)$ can be transmitted over the multiple access channel to recover the function $f(U_1, U_2)$ with distortion $D$ if for any $\epsilon > 0$ there is an $n_0$ such that for all $n > n_0$ there exist encoders $f_{E,i}^n : \mathcal{U}_i^n \times \mathcal{Z}_i^n \to \mathcal{X}_i^n$, $i = 1, 2$ and a decoder $f_D^n : \mathcal{Y}^n \times \mathcal{Z}^n \to \hat{\mathcal{G}}^n$ such that $\frac{1}{n} E \left[ \sum_{j=1}^n d(f(U_{1j}, U_{2j}), \hat{G}_j) \right] \leq D + \epsilon$ where $\hat{G}^n = f_D(Y^n, Z^n)$ and $\mathcal{U}_i, \mathcal{Z}_i, \mathcal{Z}, \mathcal{X}_i, \mathcal{Y}, \hat{\mathcal{G}}$ are the sets in which $U_i, Z_i, Z, X_i, Y, \hat{G}$ take values.

We denote the joint distribution of $(U_1, U_2)$ by $p(u_1, u_2)$. Since the MAC is memoryless, $p(y^n|x_1^n, x_2^n) = \prod_{j=1}^n p(y_j|x_{1j}, x_{2j})$. $X \leftrightarrow Y \leftrightarrow Z$ will indicate that $\{X, Y, Z\}$ form a Markov chain.

Now we state the main theorem.

*Theorem 1:* A function $f$ of the sources $(U_1, U_2)$ can be transmitted over the multiple access channel with distortion $D$ if there exist random variables $(W_1, W_2, X_1, X_2)$ such that

(1) $p(u_1, u_2, z_1, z_2, z, w_1, w_2, x_1, x_2, y) = p(u_1, u_2, z_1, z_2, z)$
$p(w_1|u_1, z_1)p(w_2|u_2, z_2)p(x_1|w_1)p(x_2|w_2)p(y|x_1, x_2)$

and

(2) there exists a function $f_D : \mathcal{W}_1 \times \mathcal{W}_2 \times \mathcal{Z} \to \hat{\mathcal{G}}$ such that $E[d(G, \hat{G})] \leq D$, where $G = f(U_1, U_2)$, $\hat{G} = f_D(W_1, W_2, Z)$ and the constraints

$$I(U_1, Z_1; W_1|W_2, Z) < I(X_1; Y|X_2, W_2, Z),$$
$$I(U_2, Z_2; W_2|W_1, Z) < I(X_2; Y|X_1, W_1, Z), \quad (1)$$
$$I(U_1, U_2, Z_1, Z_2; W_1, W_2|Z) < I(X_1, X_2; Y|Z)$$

are satisfied where $\mathcal{W}_i$ are the sets in which $W_i$ take values.

Proof of Theorem 1 is outlined in the Appendix. In the proof of Theorem 1 the encoding scheme involves finding $(W_1^n, W_2^n)$ from the sources $(U_1^n, U_2^n)$ and the side information $Z_1^n, Z_2^n$ by exploiting the structure of the function and the distortion permitted, followed by a correlation preserving mapping to the channel codewords $(X_1^n, X_2^n)$. The correlation required should be such that the inequalities in (1) are satisfied. The decoding approach involves first decoding $(W_1^n, W_2^n)$ and then obtaining the estimate $\hat{G}$ as a function of $(W_1^n, W_2^n)$ and the decoder side information $Z^n$.

One of the problems in applying Theorem 1 is that the auxiliary random variable $W$'s defined are not explicit enough. We propose good joint source-channel coding schemes for obtaining $W$'s in the subsequent sections.

## III. SPECIAL CASES

In the following we show that our result contains several previous studies as special cases.

### A. Lossy coding of functions with side information

Choose $Y = (X_1, X_2)$ in (1) and also $X_i = W_i$, $i = 1, 2$. Then the R.H.S of (1) evaluate to $H(W_1|W_2), H(W_2|W_1)$ and

$H(W_1, W_2)$ respectively. Thus we get a rate region

$$\begin{align}
R_1 &> I(U_1, Z_1; W_1 | W_2, Z), \\
R_2 &> I(U_2, Z_2; W_2 | W_1, Z) \qquad (2)\\
R_1 + R_2 &> I(U_1, U_2, Z_1, Z_2; W_1, W_2 | Z).
\end{align}$$

If we take $(Z_1, Z_2) \perp (U_1, U_2, Z)$ and $U_2 = W_2 =$ constant, then $R_1 > I(U_1; W_1|Z)$ and we have a decoder such that from $W_1$ and $Z$ it can estimate $f(U_1, Z)$ within distortion $D$. This recovers the result in [28].

*B. Lossy coding of functions with side information and remote sources*

The results of [11] can be recovered by applying the theorem to $\tilde{U}_1$ and $\tilde{Z}$ where $\tilde{U}_1$ and $\tilde{Z}$ are noisy versions of $U_1$ and $Z$. Similar to the above example we need $R_1 > I(\tilde{U}_1; W_1|\tilde{Z})$. We choose the distortion measure to allow dependence of the distortion measure on $Z$. So from the theorem we need $E[\tilde{d}(\tilde{U}_1, \tilde{Z}, g(W, \tilde{Z}))] \le D$ where $\tilde{d}(\tilde{U}_1, \tilde{Z}, g(W, \tilde{Z})) = \sum_{(u_1 \times z) \in \mathcal{U}_1 \times \mathcal{Z}} p(u_1, z|\tilde{u}_1, \tilde{z}) d(f(u_1, z), g(w_1, \tilde{z}), z)$.

*C. Distributed joint source-channel coding over a multiple access channel*

If we take $f(U_1, U_2) = (U_1, U_2)$, then we recover the results in [24] for joint source-channel coding of correlated sources over a multiple access channel with side information.

## IV. JOINT-SOURCE CHANNEL CODING FOR DISCRETE SOURCES AND CHANNELS

In this Section we extend the graph coloring technique given in [9] to joint source-channel coding of functions of discrete sources. The channel alphabets can be discrete or continuous (in particular the channel can be a GMAC). We explain the graph related terminologies through relevant previous work.

Consider sources $(U_1, U_2) \in (\mathcal{U}_1, \mathcal{U}_2)$. The following definitions are from [25]. The *characteristic graph* $G = (V, E)$ of $U_1$ w.r.t. $U_2$ and $f$ has vertex set $V = \mathcal{U}_1$ and the edge set $E$ consisting of edges $(u_{11}, u_{12}) \in \mathcal{U}_1^2$ if there exists a $u_2 \in \mathcal{U}_2$ such that $p(u_{11}, u_2) p(u_{12}, u_2) > 0$ and $f(u_{11}, u_2) \ne f(u_{12}, u_2)$. Hence $G$ represents the confusability graph from the decoder's perspective. The *conditional graph entropy* is defined as

$$H_G(U_1|U_2) = min_{U_1 \in W \in \Gamma(G)} I(W; U_1|U_2),$$

where $W \leftrightarrow U_1 \leftrightarrow U_2$ and $\Gamma(G)$ denotes the set of stable sets of $G$. [1]

The vertex coloring of a graph $G = (V, E)$ is any function $c : V \to \mathcal{N}$ where $\mathcal{N}$ is the set of natural numbers, such that $(u_{11}, u_{12}) \in E$ implies $c(u_{11}) \ne c(u_{12})$. The entropy of the coloring is the entropy induced by the distribution of $(U_1, U_2)$ on the colors and is called *chromatic entropy*.

The *OR-product graph* ([1]) of $G$ is denoted by $G^n = (V_{OR}, E_{OR})$ where $V_{OR} = V^n$ and two vertices $(\mathbf{u_{11}}, \mathbf{u_{12}}) \in E_{OR}$ if any of its components $(u_{11i}, u_{12i}) \in E$.

[1] A stable set is a set of vertices where no two of its elements are adjacent.

The *conditional chromatic entropy* $H_G^\xi(U_1) = min_{G-colorings\ c} H(c(U_1)|U_2)$ is defined in [8] and it is proved that $lim_{n \to \infty} \frac{1}{n} H_{G^n}^\xi(\mathbf{U_1}|\mathbf{U_2}) \to H_G(U_1|U_2)$. This result is an extension of Korner's result in [13] and implies that for large $n$ we can color most of the OR-product graph $G^n$ (the component of the graph $G^n$ left uncolored has low probability) and send the colors to achieve an optimal coding scheme. Thus the graph coloring scheme can decouple the functional coding from the correlation coding.

In [9] the achievability results are shown for the distributed functional source coding problem provided the joint distribution satisfies a zigzag condition, i.e., $p(x_1, y_1) > 0$ and $p(x_2, y_2) > 0$ imply either $p(x_1, y_2) > 0$ or $p(x_2, y_1) > 0$. The achievable rate region is obtained by first coloring the OR-product graph and then compressing the colors through a Slepian-Wolf coding scheme.

Our coding scheme also relies upon coloring the graph and thus separating functional source coding and the correlation coding. We assume that the zigzag condition holds. If we allow distortion in evaluating the function we can color a subgraph of the original graph [7] using a choice of colors that achieve the minimum chromatic entropy. Once we color the graphs at both encoders, the correlated colors are sent to the decoder using good joint source-channel coding schemes. The joint source-channel codes preserve the correlation in the colors and thus help in combating channel noise. The effectiveness of joint source-channel codes is shown in [21] and [24]. Using joint source-channel coding schemes the colors are losslessly recovered at the decoder and the zigzag condition ensures that the function can be recovered within a given distortion $D$. We demonstrate the performance of this joint source-channel code for functions through an example where we show that this joint source-channel coding scheme outperforms the scheme with Slepian-Wolf coding on colored outputs.

## V. EXAMPLE

Let $(U_1, U_2)$ take values in the set $\{1, 2, 3\}$ and be jointly distributed with $p(u_1, u_2) = 1/6$ for all $u_1 \ne u_2$. The function to be computed at the receiver is $f(u_1, u_2) = 1$ if $u_1 > u_2$ and 0 otherwise. The zig-zag condition holds for this case. Let the sources be transmitted over a multiple access channel with input alphabets $\{0, 1\}$ and the output alphabet $\{0, 1, 2\}$. The MAC output $Y = X_1 + X_2$. The sum capacity of this channel is 1.5 bits.

If we consider transmitting the sources losslessy without exploiting the correlation and the structure of the function we need $log(3) = 1.58$ bits at each encoder. Hence the sum rate is 3.16 bits and $(U_1, U_2)$ cannot be transmitted over the channel.

Next we exploit the correlation between the sources and not the structure of the function, i.e., we recover the sources at the decoder and then reconstruct the function. We compress the sources using Slepian-Wolf codes with sum rate of $H(U_1, U_2) = 2.58\ bits$. Still we will not be able to transmit over the channel with independent code words.

The characteristic graph defined on the set $\{1, 2, 3\}$ at encoder 1 with respect to $f$ and $U_2$ will have a single edge between 1 and 3. Similarly for encoder 2. Now we consider coloring of the graph. Let $C_1(1) = C_1(2) = 0$ and $C_1(3) = 1$ at encoder 1. Similarly define $C_2$ for the second encoder. The coloring takes into account the structure of the function. If we don't exploit the correlation we can send the colors at 0.918 bits at each encoder. This requires a sum rate of 1.8366. One sees that exploiting the structure of the function substantially reduces the data rate. However the channel still does not support the transmission.

Next we do Slepian-Wolf coding on the colors and then use independent channel code words. The joint entropy of the colors $H(C_1, C_2) = 1.58$ $bits$ and the channel capacity is 1.5 bits. Hence even in this case the function can not be evaluated losslessly.

Next we consider a joint source-channel code where the colors are mapped to channel alphabets $X_1 = C_1$ and $X_2 = 1 - C_2$. This produces correlated channel codewords and it meets the condition for joint source-channel coding since $H(C_1, C_2) = I(X_1, X_2; Y) = 1.58$ $bits$ and hence the colors can be recovered losslessly at the decoder. Given the colors the function can be evaluated. Thus we notice that we can exploit the correlation between the sources, the structure of the function and the structure of the channel to construct joint source-channel coding schemes via graph coloring.

Finally we show the use of side information. The decoder is provided with $Z = |U_1 - U_2|$. The colors are compressed via Slepian-Wolf taking into consideration the side information $Z$ also. Hence the sum rate required is reduced to $1.32$ $bits$ and we can transmit the sources losslessy via independent channel codes.

## VI. Joint source channel coding for continuous sources and channels

In this section we consider joint source-channel coding for continuous alphabet sources.

Let $f$ be Lipschitz: there is an $\alpha > 0$ such that $|f(u_1, u_2) - f(u'_1, u'_2)| < \alpha d((u_1, u_2), (u'_1, u'_2))$ for all $(u_1, u_2)$, $(u'_1, u'_2)$. Then $E[|f(U_1, U_2) - f(\hat{U}_1, \hat{U}_2)|] < \alpha E[d((U_1, U_2), (\hat{U}_1, \hat{U}_2))]$ where $(\hat{U}_1, \hat{U}_2)$ is an estimate of $(U_1, U_2)$ at the decoder. Thus for distortion within $D$ in function estimation, if $E[d((U_1, U_2), (\hat{U}_1, \hat{U}_2))] < \delta$ where $\delta = D/\alpha$, our requirements on function estimation are satisfied. This motivates us to obtain the following schemes. Of course these schemes can be used in conjunction with other properties of $f$ (instead of Lipschitz as mentioned above).

### A. Coding of $U_i$, $i = 1, 2$

**Scheme 1**: Obtain $(W_1^n, W_2^n)$ via vector quantization of $(U_1^n, U_2^n)$ (with distortion $\leq \delta$). Transmit $(W_1^n, W_2^n)$ losslessly via a joint source-channel coding scheme. Approximate $f(U_1, U_2)$ at the decoder via $f(W_1, W_2)$. This scheme exploits only the Lipschitz property of the function.

**Scheme 2**: Obtain $(W_1^n, W_2^n)$ via vector quantization as in scheme 1. Apply graph coloring to $(W_1^n, W_2^n)$ as in Section IV. These colors are further compressed via Slepian-Wolf coding. The compressed colors are sent over the MAC using independent channel code words. This scheme corresponds to the separation based (SB) scheme ([18]) for function coding.

**Scheme 3**: Obtain $(W_1^n, W_2^n)$ and apply graph coloring as in scheme 2. The colors are sent over the MAC using a joint source-channel code and the functions are estimated from the colors.

We compare the performance of the three schemes via examples. In all our examples below we see that scheme 3 performs the best although sometimes scheme 1 and scheme 3 provide the same outcome. In the following we not only consider Lipschitz functions but also others to illustrate that these ideas have general applicability.

### B. Difference of Gaussian sources over a Gaussian MAC

We consider correlated Gaussian sources $(U_1, U_2)$ with mean zero, variance $\sigma^2$ and positive correlation $\rho$ and are interested in estimating $U_1 - U_2$ at the decoder. Each of the channel inputs is power constrained to $P$ and the sources are sent over a GMAC with receiver noise variance $= 1$. This function has been considered in [15] and [26].

We compare the performance of the above schemes along with a few other schemes natural for this setting. Some of these schemes have been already studied in literature. Of course as against these schemes the three schemes proposed above are applicable to general classes of functions, sources and channels.

To put the performance of different coding schemes in a proper perspective, we also consider a lower bound on achievable distortion by considering a centralized encoder which has both the sources. The channel input $X = \alpha(U_1 - U_2)$ is power constrained to $2P$. Hence it is equivalent to a point to point channel given by $Y = X + V$ where $V$ is independent Gaussian noise with zero mean and variance 1. The lower bound obtained via this system on the mean square distortion (MSE) of the function is given by

$$D_{cen} = \frac{2\sigma^2(1-\rho)}{1+2P}. \quad (3)$$

We also study a scheme where $X_1 = \sqrt{P}U_1$ and $X_2 = -\sqrt{P}U_2$. In this scheme the channel codewords are scaled versions of the source. Thus we call it Amplify and Forward (AF). At the decoder the difference of the sources are estimated from the channel output as $E[(U_1 - U_2)|Y]$. Then the MSE is

$$D_{AF} = \frac{2\sigma^2(1-\rho)}{1+2P(1-\rho)}. \quad (4)$$

From (3) and (4) we see that $D_{AF}$ approaches $D_{cen}$ at low $\rho$. A similar conclusion holds if $f$ is any additively separable function.

Lattice coding schemes in [10] have been shown to improve upon the AF, discussed above for this system. In addition to considering these schemes we also compare performance with the lattice coding scheme given in [26].

Next we consider Scheme 2. For the specific case of difference of Gaussian sources over a GMAC, the characteristic graph formed from $(W_1, W_2)$ and $f$ is complete. Hence

$H_G(X) = H(X)$ and coloring the graph does not give any advantage. The colors compressed via Slepian-Wolf are sent over the MAC using independent channel code words.

Next consider Scheme 1. In the example the colors are mapped to Gaussian codewords, using the correlation preserving mapping provided in [17]. This scheme is shown to perform well in [22]. At the decoder the function is estimated from the colors. Because the graph is complete Scheme 3 is same as Scheme 1.

We compare the schemes discussed above in Figs. 1 and 2 for $\rho = 0.5$ and $0.75$. Since Lattice codes need $P > 0.5$ for error free decoding, we compare the schemes only in this region. From the figures we see that Scheme 1 performs better than scheme 2 for all SNR and $\rho$. Also for lower $\rho$, AF is closer to the lower bound than the lattice coding scheme. In this example, since the function was matched to the GMAC, AF and lattice codes provide considerable gains. Next we show by an example that this is not always true.

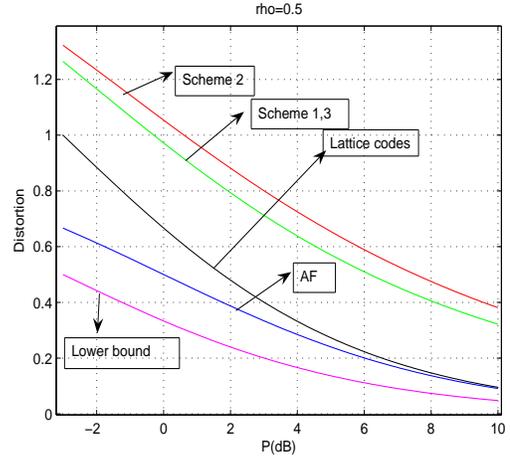

Fig. 2. Comparison of schemes for joint source-channel coding of functions, $\rho = 0.75$

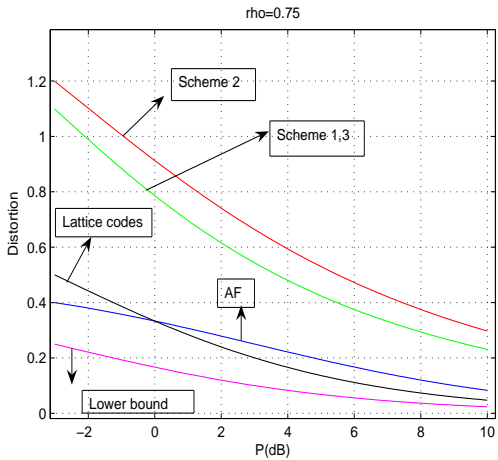

Fig. 1. Comparison of schemes for joint source-channel coding of functions, $\rho = 0.5$

### C. Binary function of Gaussian sources over a Gaussian MAC

Consider the sources and the channel to be the same as in the above example. The function to be computed is

$$f(min(U_1, U_2) > 0) = 1, \ f(min(U_1, U_2) \leq 0) = 0.$$

The natural $W_i's$ for this case are as follows. The $W_i$ at encoder $i$ is $= 1$, if $U_i > 0$ else $W_i = 0$. The $(W_1, W_2)$ thus obtained are jointly distributed with zero mean and covariance

$$\begin{pmatrix} 1/4 + (sin^{-1}\rho)/2\pi & (cos^{-1}\rho)/2\pi \\ (cos^{-1}\rho)/2\pi & 1/4 + (sin^{-1}\rho)/2\pi \end{pmatrix}, \quad (5)$$

where $\rho$ is the correlation between $(U_1, U_2)$.

We color the $W_i's$ at each encoder. Since the characteristic graph is complete, coloring does not give any advantage. For $\rho = 0.75$ and $P = 5$, by Slepian-Wolf coding $(W_1, W_2)$ can be compressed to $H(W_1, W_2) = 1.778 \ bits$ (this scheme is similar to scheme 2). However the channel capacity is only 1.729 $bits$ and $(W_1, W_2)$ cannot be reliably transmitted. Thus the function cannot be losslessly computed.

Next we send $(W_1, W_2)$ using a joint-source channel coding scheme (similar to scheme 1 and 3). Such bit to Gaussian mapping schemes are discussed in [21]. The correlation between $(W_1, W_2)$ is 0.54 and we consider mappings of $(W_1, W_2)$ to jointly Gaussian codewords with correlation 0.3. Then the channel supports sum rates till 1.903 bits and hence the function can be losslessly computed.

If we consider AF or the lattice coding scheme, the sources cannot be recovered losslessly at the decoder (since $P$ is finite) and hence the function also cannot be losslessly computed with these schemes.

### D. Uniform continuous sources over a discrete MAC

So far we have seen that graph coloring does not help for continuous sources. But the next example shows that it can indeed still be useful. $(U_1, U_2)$ take values in $[0, \ 1] \times [0, \ 1]$. The joint density is 0 in $[0 \ \ 1/3] \times [0 \ \ 1/3]$, $[1/3 \ \ 2/3] \times [1/3 \ \ 2/3]$ and $[2/3 \ \ 1] \times [2/3 \ \ 1]$. In the rest of the regions the joint density is uniform. The sources are to be transmitted through the MAC $(Y = X_1 + X_2)$, given in the example in Section V, with a specified distortion. The distortion measure $d(x, x') = |x - x'|$. The decoder is interested in finding $|U_1 - U_2|$ within a distortion of $16.67\%$.

We quantize $(U_1, U_2)$ via a uniform grid on $[0 \ 1] \times [0 \ 1]$ dividing each axis into three parts. Hence there are 9 representation points. These representation points are at the center of each square cell. Thus each $W_i$, $i = 1, 2$ takes 3 values. Graph coloring is applied to $W_i's$. The characteristic graph of $W_1$ has three vertices $\{1, 2, 3\}$ and there is an edge between $1 - 2$ and $2 - 3$. Hence we can color this using two colors. If we use Slepian-Wolf coding on the colors, they can be compressed to 1.58 bits whereas the channel capacity is only 1.5 bits. Hence scheme 2 will not be able to support this transmission and the distortion constraint is not met. But these colors can be jointly transmitted across the channel by using the joint source-channel coding scheme: $X_1 = C_1$, $X_2 = 1 - C_2$. Thus we can transmit with a distortion of 16.67 % via scheme 3.

If we try to send $(W_1, W_2)$ without coloring (i.e., not exploiting the structure of the function) using a joint source channel code, $H(W_1, W_2) = 2.58\ bits$. The channel capacity is only $1.5\ bits$. Thus we cannot transmit $(W_1, W_2)$ losslessly via scheme 1 and again distortion constraint is not met.

If we use vector quantizers instead of scalar quantizers as in this example the distortion is further reduced in the coloring scheme.

## VII. Conclusions and Discussion

In this paper we provide sufficient conditions for lossy transmission of functions over a MAC. Efficient joint source-channel coding schemes are developed for transmission of discrete and continuous alphabet sources. It is shown that by exploiting the function structure substantial savings in transmission rate can be obtained. Although, a good coding scheme can depend upon the function, the source statistics and the channel, we identify schemes which perform well for various classes of functions, sources and channels.

The results in this paper are information-theoretic and hence address the fundamental limitations of the communication system. From our results one can determine if one can transmit data on a MAC so that it will be possible for the receiver to compute the value of the function within a given distortion. However, one needs to develop practical coding schemes to design such a system and the aim of the designer is to develop codes that perform close to the information theoretic limits provided. Also, our results, as often happen in information theory, can provide significant insights in designing practical codes. For example, now practically implementable codes for Slepian-Wolf coding are available ([20], [4]). Thus our Scheme 2 can be implemented in practice. Recently there is also some work on practical implementation of joint source-channel coding ([29]).

## Appendix

We show the achievability of all points in the rate region (1). The proof is an extension of the proof in [24]. Thus we will outline the proof. Initially we prove it for bounded $d$ and discrete $(U_1, U_2, Z_1, Z_2, Z)$. At the end we will generalize these assumptions.

*Proof*: Fix $p(w_1|u_1, z_1), p(w_2|u_2, z_2), p(x_1|w_1), p(x_2|w_2)$ as well as $f_D(.)$ satisfying the distortion constraints of the form $E[d(f(U_1, U_2), g(W_1, W_2, Z))] \leq D$, where $g$ is a deterministic function.

*Codebook Generation*: Let $R_i^{'} = I(U_i, Z_i; W_i) + \delta, i \in \{1, 2\}$ for some $\delta > 0$. Generate $2^{nR_i^{'}}$ codewords of length $n$, sampled iid from the marginal distribution $p(w_i), i \in \{1, 2\}$. For each $w_i^n$ independently generate sequence $X_i^n$ according to $\prod_{j=1}^n p(x_{ij}|w_{ij}), i \in \{1, 2\}$. Call these sequences $x_i(w_i^n), i \in 1, 2$. Reveal the codebooks to the encoders and the decoder.

*Encoding*: For $i \in \{1, 2\}$, given the source sequence $U_i^n$ and $Z_i^n$, the $i^{th}$ encoder looks for a codeword $W_i^n$ such that $(U_i^n, Z_i^n, W_i^n) \in T_\epsilon^n(U_i, Z_i, W_i)$ and then transmits $X_i(W_i^n)$ where $T_\epsilon^n(.)$ is the set of $\epsilon$-weakly typical sequences ([6]) of length $n$.

*Decoding*: Upon receiving $Y^n$, the decoder finds the unique $(W_1^n, W_2^n)$ pair such that $(W_1^n, W_2^n, x_1(W_1^n), x_2(W_2^n), Y^n, Z^n) \in T_\epsilon^n$. If it fails to find such a unique pair, the decoder declares an error and incurres a maximum distortion of $d_{max}$.

In the following we show that the probability of error for the encoding decoding scheme tends to zero as $n \to \infty$. The error can occur because of the following four events **E1**-**E4**. We show that $P(\mathbf{Ei}) \to 0$, for $i = 1, 2, 3, 4$.

**E1** The encoders do not find the codewords. However from rate distortion theory [6], page 356, $\lim_{n \to \infty} P(E_1) = 0$ if $R_i^{'} > I(U_i, Z_i; W_i), i \in 1, 2$.

**E2** The codewords are not jointly typical with $Z^n$. Probability of this event goes to zero from the extended Markov Lemma.

**E3** There exists another codeword $\hat{w}_1^n$ such that $(\hat{w}_1^n, W_2^n, x_1(\hat{w}_1^n), x_2(W_2^n), Y^n, Z^n) \in T_\epsilon^n$. Define $\alpha \stackrel{\Delta}{=} (\hat{w}_1^n, W_2^n, x_1(\hat{w}_1^n), x_2(W_2^n), Y^n, Z^n)$. Then,

$$P(\mathbf{E3}) = Pr\{\text{There is } \hat{w}_1^n \neq w_1^n : \alpha \in T_\epsilon^n\}$$
$$\leq \sum_{\hat{w}_1^n \neq W_1^n : (\hat{w}_1^n, W_2^n, Z^n) \in T_\epsilon^n} Pr\{\alpha \in T_\epsilon^n\} \quad (6)$$

It can be shown that P(**E3**) tends to zero if $I(U_1, Z_1; W_1|W_2, Z) < I(X_1; Y|X_2, W_2, Z)$. Similarly $I(U_2, Z_2; W_2|W_1, Z) < I(X_2; Y|X_1, W_1, Z)$.

**E4** There exist other codewords $\hat{w}_1^n$ and $\hat{w}_2^n$ such that $\alpha \stackrel{\Delta}{=} (\hat{w}_1^n, \hat{w}_2^n, x_1(\hat{w}_1^n), x_2(\hat{w}_2^n), y^n, z^n) \in T_\epsilon^n$. Then,

$$P(\mathbf{E4}) = Pr\{\text{There is } (\hat{w}_1^n, \hat{w}_2^n) \neq (w_1^n, w_2^n) : \alpha \in T_\epsilon^n\}$$
$$\leq \sum_{(\hat{w}_1^n, \hat{w}_2^n) \neq (w_1^n, w_1^n) : (\hat{w}_1^n, \hat{w}_2^n, z^n) \in T_\epsilon^n} Pr\{\alpha \in T_\epsilon^n\} (7)$$

It can be shown that P(**E4**) tends to zero if $I(U_1, U_2, Z_1, Z_2; W_1, W_2|Z) < I(X_1, X_2; Y|Z)$.

Thus as $n \to \infty$, with probability tending to 1, the decoder finds the correct sequence $(W_1^n, W_2^n)$ which is jointly weakly $\epsilon$-typical with $(U_1^n, U_2^n, Z^n)$.

The fact that $(W_1^n, W_2^n)$ are weakly $\epsilon$-typical with $(U_1^n, U_2^n, Z^n)$ does not guarantee that $f_D^n(W_1^n, W_2^n, Z^n)$ will satisfy the distortion $D$. For this, one needs that $(W_1^n, W_2^n)$ are distortion weakly $\epsilon$- typical (in the sense that $d(f(u_1^n, u_2^n), f_d^n(w_1^n, w_2^n, z^n)) \leq D$ is also satisfied by the weakly $\epsilon$-typical set) ([6]) with $(U_1^n, U_2^n, Z^n)$. Let $T_{D,\epsilon}^n$ denote the set of distortion typical sequences ([6]). Then by strong law of large numbers $P(T_{D,\epsilon}^n|T_\epsilon^n) \to 1$ as $n \to \infty$. Thus the distortion constraints are also satisfied by $(W_1^n, W_2^n)$ obtained above with a probability tending to 1 as $n \to \infty$. Therefore, if distortion measure $d$ is bounded $\lim_{n \to \infty} E[d(f(U_1^n, U_2^n)), f_D(W_1^n, W_2^n, Z^n)] \leq D + \epsilon\ i = 1, 2$.

If there exist $u_1^*, u_2^*$ such that $E[d(f(U_1^n, U_2^n)), f(u_1^{*n}, u_2^{*n})] < \infty$, then the result extends to unbounded distortion measures also as follows. Whenever the decoded $(W_1^n, W_2^n, Z^n)$ are not in the distortion typical set then we estimate $f(U_1^n, U_2^n)$ as $f(u_1^{*n}, u_2^{*n})$. Then for $i = 1, 2$, for large enough $n$,

$$E[d(f(U_1^n, U_2^n), f_D(W_1^n, W_2^n, Z^n))] \leq D_i + \epsilon$$
$$+ E[d(f(U_1^n, U_2^n), f(u_1^{*n}, u_2^{*n}))\mathbf{1}_{(T_{D,\epsilon}^n)^c}] \quad (8)$$

Since $E[d(f(U_1^n, U_2^n)), f(u_1^{*n}, u_2^{*n})] \leq \infty$ and $P[(T_{D,\epsilon}^n)^c] \to 0$ as $n \to \infty$, the last term of (8) goes to zero as $n \to \infty$.

## REFERENCES

[1] N. Alon and A. Orlitsky. Source coding and graph entropies. *IEEE Trans. Inform. Theory*, 42(5):1329–1339, Sept. 1996.
[2] J. Cardinal, S. Fiorini, and G. V. Assche. On minimum entropy graph colorings. *IEEE ISIT*, 2004.
[3] J. Cardinal, S. Fiorini, and G.Joret. Minimum entropy coloring. *Lecture notes in computer science, ser. International symposium on Algorithms and computation, Springer-Verlag*, pages 819–828, 2005.
[4] T. P. Coleman, A. H. Lee, M. Medard, and M. Effros. Low-complexity approaches to Slepian-Wolf near-lossless distributed data compression. *IEEE Trans. Inform. Theory*, 52(8):3546–3561, Aug. 2006.
[5] T. M. Cover, A. E. Gamal, and M. Salehi. Multiple access channels with arbitrarily correlated sources. *IEEE Trans. Inform. Theory*, IT - 26:648–657, 1980.
[6] T. M. Cover and J. A. Thomas. *Elements of Information theory*. Wiley Series in Telecommunication, N.Y., 2004.
[7] V. Doshi, D. Shah, and M. Medard. Source coding with distortion through graph coloring. *IEEE ISIT*, May 2007.
[8] V. Doshi, D. Shah, M. Medard, and S. jaggi. Graph coloring and conditional graph entropy. *Asilomer conference on Signals, Systems and Computers*, Nov. 2006.
[9] V. Doshi, D. Shah, M. Medard, and S. jaggi. Distributed functional compression through graph coloring. *Proceedings of data compression conference*, pages 93–102, March 2007.
[10] U. Erez, S. Litsyn, and R. Zamir. Lattices which are (almost) good for everything. *IEEE Trans. Inform. Theory*, 51(10):3401–3416, Oct. 2005.
[11] H. Feng, M. Effros, and S. Savari. Functional source coding for networks with receiver side information. *Allerton conference on communication, control and computing*, Sept. 2004.
[12] A. Giridhar and P. R. Kumar. Computing and communicating functions over sensor networks. *IEEE JSAC*, 23(4), 2005.
[13] J. Korner. Coding of an information source having ambiguous alphabet and the entropy of graphs. *Sixth Prague conference on information theory*, pages 411–425, 1973.
[14] J. Korner and K. Marton. How to encode modulo two sum of binary sources. *IEEE Trans. Inform. Theory*, 25:219–221, March 1979.
[15] D. Krithivasan and S. Pradhan. Lattices for distributed source coding: Jointly gaussian sources and reconstruction of a linear function. *Preprint available at http://arxiv.org/abs/0707.3461*, 2007.
[16] D. Krithivasan and S. S. Pradhan. An achievable rate region for distributed source coding with reconstruction of an arbitrary function of the sources. *IEEE ISIT, Toronto*, 2008.
[17] A. Lapidoth and S. Tinguely. Sending a bi- variate Gaussian source over a Gaussian MAC. *IEEE ISIT 06*, 2006.
[18] B. Nazar and M. Gastpar. Computation over multiple access channel. *IEEE Trans. Inform. Theory*, Oct. 2007.
[19] A. Orlitsky and J. R. Roche. Coding for computing. *IEEE Trans. Inform. Theory*, 47(3):903–917, March 2001.
[20] S. S. Pradhan and K. Ramchandran. Distributed source coding using syndromes DISCUS: Design and construction. *IEEE Trans. Inform. Theory*, 49(3):626 – 643, March 2003.
[21] R. Rajesh and V. Sharma. A joint source-channel coding scheme for transmission of correlated discrete sources over a Gaussian multiple access channel. *ISITA 2008, Aukland, New Zealand*.
[22] R. Rajesh and V. Sharma. Source channel coding for Gaussian sources over a Gaussian multiple access channel. *Proc. 45 Allerton conference on Computing Control and Communication*, 2007.
[23] R. Rajesh and V. Sharma. Amplify and forward for correlated data gathering over hierarchical sensor networks. *IEEE WCNC, Budapest, Hungary*, 2009.
[24] R. Rajesh, V. K. Varshneya, and V. Sharma. Distributed joint source-channel coding on a multiple access channel with side information. *IEEE ISIT 2008*.
[25] G. Simonyi. Graph entropy: A survey. *Combinatorial optimization, DIMACS series on discrete math and computer science*, 20:399–441, 1995.
[26] R. Soundararajan and S. Viswanath. Communicating difference of correlated gaussian sources over a MAC. *arxiv, Available online ; arxiv.org/pdf/0812.1091*, 2008.
[27] H. Witsenhausen. The zero error side information problem and chromatic numbers. *IEEE Trans. Inform. Theory*, 22(1):1–11, Jan. 1976.
[28] H. Yamamoto. Wyner-ziv theory for general function of correlated soruces. *IEEE Trans. Inform. Theory*, 28(5):803–807, Sept. 1982.
[29] W. Zong and J. G. Frias. LDGM codes for channel coding and joint source-channel coding of correlated sources. *EURASIP Journal on Applied Signal Processing*, pages 942–953, 2005.